\documentclass[aps,prx,twocolumn,superscriptaddress,nofootinbib,longbibliography]{revtex4-2}
\usepackage[colorlinks=true,citecolor=blue,linkcolor=blue,urlcolor=blue]{hyperref}
\usepackage{amsmath, amssymb}
\usepackage{graphicx, nicefrac, textcomp}
\usepackage{mwe}
\usepackage{braket}
\usepackage{color}
\usepackage{dcolumn}
\usepackage[utf8]{inputenc}
\usepackage[T1]{fontenc}
\usepackage{float}
\usepackage{dsfont}
\usepackage{svg}
\usepackage[version=4]{mhchem}
\usepackage[toc,page]{appendix}
\usepackage{soul}
\usepackage{algorithm}
\usepackage{algpseudocode}
\usepackage{multirow}

\newcommand{\bfr}{\mathbf{r}}

\newcommand{\hsic}{\frac{1}{2}\text{SIC}}
\newcommand{\lss}{\text{LSSIC}}

\begin{document}
\title{Locally Scaled Self-Interaction Corrected Energy Functionals with Complex Optimal Orbitals}

\author{Jukka John}
\affiliation{Science Institute and Faculty of Physical Sciences, University of Iceland, 107 Reykjav\'ik, Iceland}

\author{Hlynur Guðmundsson}
\affiliation{Science Institute and Faculty of Physical Sciences, University of Iceland, 107 Reykjav\'ik, Iceland}

\author{Iðunn Bj\"{o}rg Arnaldsd\'{o}ttir}
\affiliation{Science Institute and Faculty of Physical Sciences, University of Iceland, 107 Reykjav\'ik, Iceland}

\author{Hannes J\'{o}nsson}
\email{hj@hi.is}
\affiliation{Science Institute and Faculty of Physical Sciences, University of Iceland, 107 Reykjav\'ik, Iceland}

\author{Elvar \"{O}rn J\'{o}nsson}
\email{elvarorn@hi.is}
\affiliation{Science Institute and Faculty of Physical Sciences, University of Iceland, 107 Reykjav\'ik, Iceland}

\date{\today}

\begin{abstract}
We present
a fully variational 
locally scaled 
self-interaction corrected (SIC) 
energy functional 
using complex optimal orbitals. 
This represents an important milestone 
for fully variational SIC energy functionals, 
which have been shown to improve the 
prediction of the properties of atomic, 
molecular and 
solid state systems in general, 
in both ground and excited states. 
However, 
it depends on the system 
and property of the system 
whether it is beneficial 
to scale the SIC correction 
by a factor of one-half, 
which makes the application of SIC inconsistent.
In the limit of a single electron the SIC 
exactly cancels the self interaction error, 
but overcorrects the error in regions of high density 
where there is large overlap between occupied orbitals. 
The newly implemented local scaling function, 
$z(\mathbf{r})$, 
which is based on an iso-orbital indicator 
derived from considering 
the kinetic energy density 
in the iso-electron 
and many electron case, 
naturally scales the SIC correction 
from $0\leq z(\mathbf{r}) \leq 1$ 
in regions of high and low (isolated orbital) electron density. 
The locally scaled SIC framework 
is general and applicable to atomic, 
molecular 
and solid-state systems. 
\end{abstract}

\maketitle


\section{Introduction}


Kohn--Sham density functional theory (KS--DFT) \cite{kohn1965, kohn1999} 
is the most widely used method for 
electronic structure calculations in 
materials science, 
chemistry, 
and nanotechnology. 
Despite its success, 
it has several shortcomings. 
In KS--DFT, 
a local or semilocal approximation 
to the exchange-correlation energy of the 
electron density 
is typically applied. 
This transforms the problem of 
$N$ interacting electrons to 
a problem of three spatial variables 
which 
describe the total electron density. 
Therefore KS--DFT provides a 
feasible way to simulate systems 
with thousands of electrons. 
However, 
approximations to the true exchange-correlation energy 
introduce the so-called self interaction error (SIE) 
which comes from
the partial cancellation 
of the self-coulomb 
and self-exchange correlation 
interaction of the electron 
with itself 
in the effective potential.

Due to the SIE, 
the description of localized electronic states in 
semiconductors and insulators is problematic, ~\cite{laegsgaard2001,nolan2005,pacch2013,
Gudmundsdottir_Jonsson_Jonsson_2015,ivanov2023diamond,galynska2025}
the excitation energy of diffused (Rydberg) 
and valence electron states~\cite{gudmundsdottir2013,gudmundsdottir2014,
sigurdarson_yorick_jonsson_2023,elliselenius2024,birgisson2025} 
are poorly described, 
the relative energies 
of s- and d-electrons in 
transition metal molecules~\cite{Ivanov_Ghosh_Jonsson_Jonsson_2021,maniar2025atomic} 
are not balanced, 
leading to incorrect magnetic states, and
bond energies of
gas phase molecules are often 
greatly overestimated (e.g. 
O$_2$ which is off by 1 $e$V~\cite{Klüpfel_Klüpfel_Jonsson_2012}) 
and absorption energy of 
ad-molecules and intermediates 
on solid surfaces 
in catalytic reactions~\cite{Urrego-Ortiz_Builes_Illas_Calle-Vallejo_2024} 
are too inaccurate for a quantitative prediction. 


Perdew and Zunger proposed a 
procedure where the orbital-by-orbital SIE is subtracted from the total energy functional\cite{perdew1981}.  
The resulting 
Perdew-Zunger self-interaction correction (PZ-SIC) 
energy functional tends to 
overcorrect the energetics of the base functional, 
and scaling down the correction 
is necessary for a 
better estimate of 
orbital energies.~\cite{bylaska2004self,jonsson2011,dabo2010,vydrov2006}  
Empirical evidence suggests 
to include a 
scaling factor of $\frac{1}{2}$ ($\hsic$) 
which was shown 
to work well for different types 
of systems and properties~\cite{jonsson2011,Gudmundsdottir_Jonsson_Jonsson_2015,klupfel2012}. 
$\hsic$ has been applied to a wide range 
of systems, 
giving better agreement than local or semi-local KS-DFT energy functionals in calculations 
of the atomization energy of molecules,~\cite{Jonsson_Lehtola_Jonsson_2015,Lehtola_Jonsson_Jonsson_2016} 
activation energy of chemical reactions,~\cite{Klüpfel_Klüpfel_Jonsson_2012} 
electronic excitation energies~\cite{Schmerwitz_Ivanov_Jonsson_Jonsson_Levi_2022} 
and band gaps of solids.~\cite{Gudmundsdottir_Jonsson_Jonsson_2015}
This shows that the method has 
significantly improved accuracy 
over the local or semi-local KS 
functionals which are, for example,
currently used in
the vast majority of 
heterogeneous catalysis calculations.

However, this is not satisfactory in many respects; 
for example 
the correct $-1/r$ dependence 
on the effective potential 
is not captured 
if the scaling is 
lower than 1. 
Therefore, 
in order to accurately describe 
Rydberg excited states, 
it is necessary 
to scale the correction by a factor of 1 
to recover the correct $-1/r$ dependence,
and PZ-SIC has been shown 
to describe such excited states 
accurately~\cite{gudmundsdottir2013,gudmundsdottir2014,cheng2014,birgisson2025}. 
Furthermore, 
in the context of 
a localized and delocalized 
structure dependent 
charge states of 
N,N'-dimethylpiperazine (DMP), 
which have been experimentally verified, 
it has been shown that when full SIC is applied to the PBE 
functional, an energy surface analogous to hybrid functionals with a 
Fock-Exchange weight of $\frac{1}{2}$ is obtained, while the scaling of SIC by 
$\frac{1}{2}$, only produces the more delocalized charge. \cite{birgisson2025}

KS-DFT depends only on the electron density 
and individual orbitals have no physical meaning. 
The energy is invariant under unitary rotations 
of occupied orbitals. 
PZ-SIC removes the SIE orbital-by-orbital and 
in a fully variational implementation, where 
the energy is minimized with respect to all 
electronic degrees of freedom ($dE/d\psi_k^* = 0$), 
the energy depends on individual orbitals which
breaks the symmetry of the underlying 
theory and the SIC energy functional becomes 
unitary variant. Due to the full flexibility of 
fully variational SIC the lowest possible energy 
with respect to all degrees of freedom can be found,
and even greater flexibility is achieved by using complex optimal orbitals.~\cite{Lehtola:2014,Lehtola_Jonsson_Jonsson_2016}
The unitary invariance of KS-DFT can be recovered in SIC 
energy functionals by restricting 
the degrees of freedom (and therefore narrow the search space), 
such as in the Fermi-Löwdin orbital SIC method, FLOSIC \cite{Jackson2019FLOSIC,Pederson2014FermiDerivatives,LLSICORIG},
where the electron density is decomposed into 
so-called Fermi-Orbitals (FOs), and instead the energy is minimized
with respect to anchoring points in space which center the individual 
FOs.


In the present project, 
we further develop the fully variational 
SIC method and introduce
a general local scaling function which bridges the gap in the application
of fully variational SIC energy functionals 
to ground and excited states. 
In particular, 
this work lays important theoretical 
and numerical groundwork for 
improving the predictive accuracy 
of DFT in materials where accurate calculation are required for atomic or molecular properties, in combination with properties of solid substrates, such as in heterogeneous catalysis.
We implement and evaluate a 
\emph{locally scaled self-interaction correction} ($\lss$) 
scheme in the 
open source and python based 
GPAW code~\cite{GPAW,GPAW2,GPAW2024}, 
where the LSSIC includes a spatial correction based 
on the degree of orbital localization, while taking into account that the orbitals are complex numbered.
This approach retains the benefits of SIC 
in regions where it is most needed and 
accurately applied, such as for 
localised orbitals, while reducing or 
eliminating the correction in delocalised 
regions where error cancellations lead to an 
overcorrection.

\section{Theory}

In Kohn-Sham 
density functional theory~\cite{hohenberg1964,kohn1965} 
the energy of an 
electronic system is given by:
\begin{equation}
E^\mathrm{KS} = T_s + V_\mathrm{ext}[n] + E_\mathrm{C}[n]  + E_{xc}[n_{\uparrow},n_{\downarrow}]
\end{equation}
Here, 
$T_{s}$ is the kinetic energy 
of the non-interacting electrons 
whose total density corresponds 
to the ground-state density of the 
interacting electrons. 
$V_\mathrm{ext}[n]$ is 
the external potential describing attractive
electron-nuclei Coulomb interaction. 
$E_\mathrm{C}[n]$ is the repulsive
electron-electron Coulomb interaction. Finally,
$E_{xc}[n_{\uparrow},n_{\downarrow}]$ is 
the exchange-correlation energy, 
a functional form which is not known exactly 
and is approximated in practice. 
For any semi-local functional 
-- and for most hybrid functionals where fractional exact exchange is included --
the electronic system 
possesses a spurious 
self-interaction error (SIE) 
due to the fact that 
there is only partial cancellation 
of the self Coulomb and exchange energies 
in the one-electron limit.

In PZ-SIC, the SIE is corrected 
orbital-by-orbital 
by removing the self Coulomb and self exchange-correlation 
energy resulting in
\begin{equation}
E^\mathrm{SIC} = E^\mathrm{KS}[n] - a\sum_{k\sigma} \left( E_\mathrm{C}[n_{k\sigma}] +  E_{xc}[n_{k\sigma}, 0]\right)
\label{eq:SIC}
\end{equation}
where $a$ is a global scaling factor (i.e. $\frac{1}{2}$ or 1), 
and the sum is over all occupied orbitals $k$ 
in spin channel $\sigma$. 
For the locally scaled SIC ($\lss$) 
the energy functional is 
further modified 
such that the orbital-by-orbital correction 
now includes a local function 
instead of the global scaling factor,
\begin{align}
E^\mathrm{SIC} = &E^\mathrm{KS}[n] \nonumber \\
&- \sum_{k\sigma} \left( E_\mathrm{C}[n_{k\sigma},z_\sigma[\{n_{k\sigma}\}]] +  E_{xc}[n_{k\sigma}, 0, z_\sigma[\{n_{k\sigma}\}]]\right)
\end{align}
where $z_\sigma[\{n_{k\sigma}\}] = z_\sigma(\bfr)$. This 
is some local function of the 
occupied orbital spin densities. 

We adopt 
an iso-orbital indicator 
as the scaling function 
-- first introduced in the context of FLOSIC by Zope et. al.~\cite{LLSICORIG} -- 
and apply it in this work to the fully varional SIC using complex optimal orbitals. 
In its most simple form it reads: 

\begin{equation}
z_\sigma(\mathbf{r}) = \frac{\tau^W_\sigma(\mathbf{r})}{\tau_\sigma(\mathbf{r})} = \frac{\frac{|\nabla n_\sigma|^2}{8n_\sigma}}{\frac{1}{2}\sum_k|\nabla \psi_{k\sigma}|^2}
\label{eq:partial}
\end{equation}
where the boundary conditions $0\leq z_\sigma(\mathbf{r}) \leq 1$ 
are respected by construction, 
that is the local scaling function 
becomes zero in the limit of 
fully-delocalized states 
(i.e. the homogeneous electron gas limit) 
and is equal to one 
for single orbital densities (one electron systems)
or orbital densities which are isolated 
-- i.e. in regions where there is no overlap 
with other orbital densities 
(see Appendix~\ref{Derivation} for derivation). 
Here $\tau_\sigma$ is the 
positive kinetic energy density 
for occupied orbitals with spin 
$\sigma$ and 
$\tau^W_\sigma =\frac{|\nabla n_\sigma|^2}{8n_\sigma}$, 
is the von Weizs\"acker kinetic energy density \cite{ParrYang1989}. 

However, we re-derive the expressions 
for the von Weizs\"acker kinetic energy density
and Kohn-Sham kinetic energy density, taking into 
account that we have complex optimal orbitals, 
we arrive at the following generalized form of 
the iso-orbital indicator (see Appendix \ref{Derivation}):
\begin{equation}
        {z}_\sigma(\mathbf{r})
=
\frac{
\frac{|\nabla {n}_\sigma(\mathbf{r})|^2}{8 {n}_\sigma(\mathbf{r})} +
\frac{1}{2} {n}_\sigma(\mathbf{r}) \, |\nabla \theta(\mathbf{r})|^2 -
\frac{1}{4} \nabla^2 {n}_\sigma(\mathbf{r})
}{
\frac{1}{2} \sum_k f_{k\sigma} |\nabla {\psi}_{k\sigma}(\mathbf{r})|^2 -
\frac{1}{4} \nabla^2 {n}_\sigma(\mathbf{r})
}
\label{eq:total}
\end{equation}
where the second term in the 
nominator is the 
gradient of the phase term 
due to the use of complex 
numbered orbitals, 
and the third term in the 
nominator (and the second term in the denominator) 
is the Laplacian term.
The Laplacian term arises for real 
and complex valued orbitals but is
omitted in this work -- and 
instead we write the local scaling function as
\begin{equation}
z_\sigma(\mathbf{r}) = \frac{\frac{|\nabla n_\sigma(\bfr)|^2}{8n_\sigma(\bfr)} + \frac{\mathbf{J}_\sigma(\bfr)\cdot \mathbf{J}_\sigma(\bfr)}{2n_{\sigma}(\bfr)}}{\frac{1}{2}\sum_k|\nabla \psi_{k\sigma}(\bfr)|^2}
\label{eq:actual}
\end{equation}
where the gradient of the complex phase term is written in terms of 
the current density $\mathbf{J}_\sigma(\bfr) = -\frac{i}{2}\sum_k\left[\psi_{k\sigma}^*(\bfr)\nabla\psi_{k\sigma}(\bfr) - \psi_{k\sigma}(\bfr)\nabla\psi^*_{k\sigma}(\bfr)\right]$, see~\cite{becke1996current,johnson2007density}.
This iso-orbital scaling function is 
consistent with the derivation 
of both energy terms with complex optimal orbitals 
and has the same 
desirable property of $0\leq z(\mathbf{r})\leq 1$.

The orbital dependent SIE terms become
\begin{equation}
 E_\mathrm{C}[n_{i\sigma},z_\sigma[n_{i\sigma}]] = \int\int \frac{z_\sigma(\bfr)n_{i\sigma}(\bfr)n_{i\sigma}(\bfr')}{|\bfr - \bfr'|}d\bfr d\bfr'
\end{equation}
and
\begin{equation}
 E_\mathrm{xc}[n_{i\sigma},0,z_\sigma[n_{i\sigma}]] = \int z_\sigma(\bfr)n_{i\sigma}(\bfr)\epsilon_\mathrm{xc}(\bfr)d\bfr
\end{equation}
where $\epsilon_\mathrm{xc}(\bfr)$ is the 
exchange-correlation energy density. 
The functional derivative with respect 
to the occupied orbitals results in 
additional terms 
for the variational minimization of 
the total SIC energy functions, 
see Appendix~\ref{Implementation}.

\begin{figure}[!tbp]
\centering
\includegraphics[width=.99\linewidth]{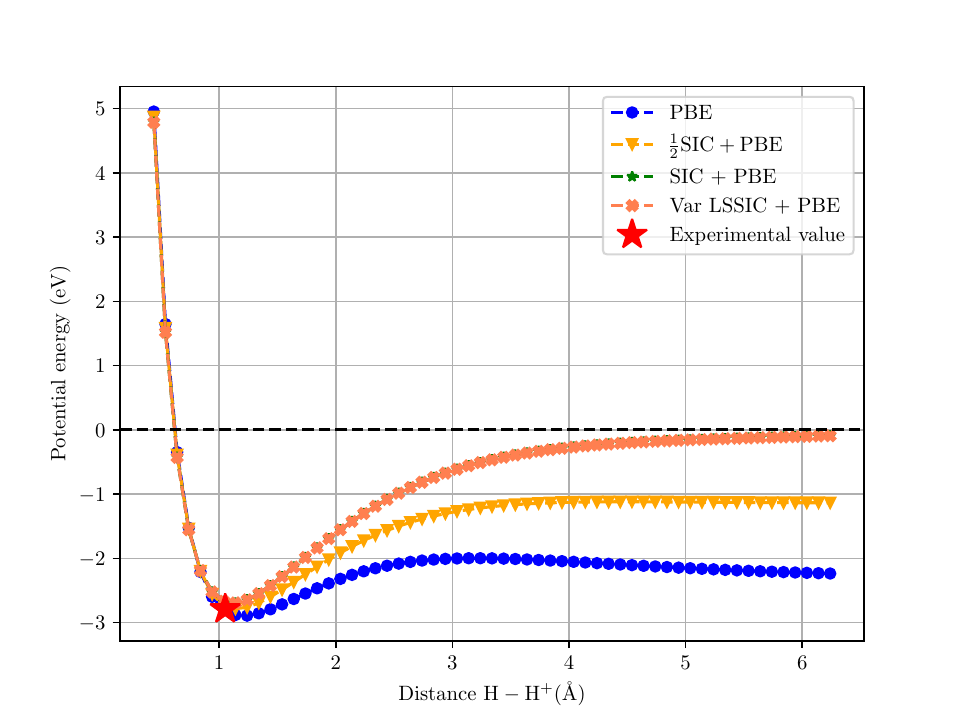}
\caption{Binding energy of H$^+_2$ 
calculated with $\hsic$ and SIC (yellow and green) 
and $\lss$ (orange). 
The base functional is PBE (blue). 
The zero of energy corresponds to 
the dissociation limit H+H$^+$. 
Note that the SIC (green) and $\lss$ (orange) curves perfectly overlap, 
since they are equivalent in the one-electron limit. The experimental estimate is marked on the plot (red star).} 
\label{fig:H2p_BE}
\end{figure}

\section{Implementation and Computations}
The locally scaled PZ-SIC energy functional is 
implemented in the open source grid-based 
projector augmented wave code GPAW \cite{GPAW1,GPAW2,GPAW2024}. 
The minimum (for ground state calculations) 
or saddle point (for excited state calculations) 
is found via a direct optimization method 
where the optimal orbitals are found 
variationally and self-consistently \cite{Ivanov_Jonsson_Vegge_Jonsson_2021,ivanov2021levijonssonjonsson,Schmerwitz_Ivanov_Jonsson_Jonsson_Levi_2022}. 
The unitary variant~\cite{ivanov2024UPAW} 
of the projector augmented wave (PAW) 
method \cite{paw1,paw2} is used 
to treat the electrons near the nuclei 
(see Appendix~\ref{Implementation}).
The core electrons for each atom 
are frozen to the result of a spherically symmetric 
calculation of the isolated atom. 
The smooth pseudo wave functions for 
the occupied MOs containing valence electrons 
are described here using a real-space grid, 
with a grid spacing of 0.16 \AA. 
In the case of SIC calculation the
wavefunctions are complex numbered.

\begin{figure}[!tbp]
\centering
\includegraphics[width=.99\linewidth]{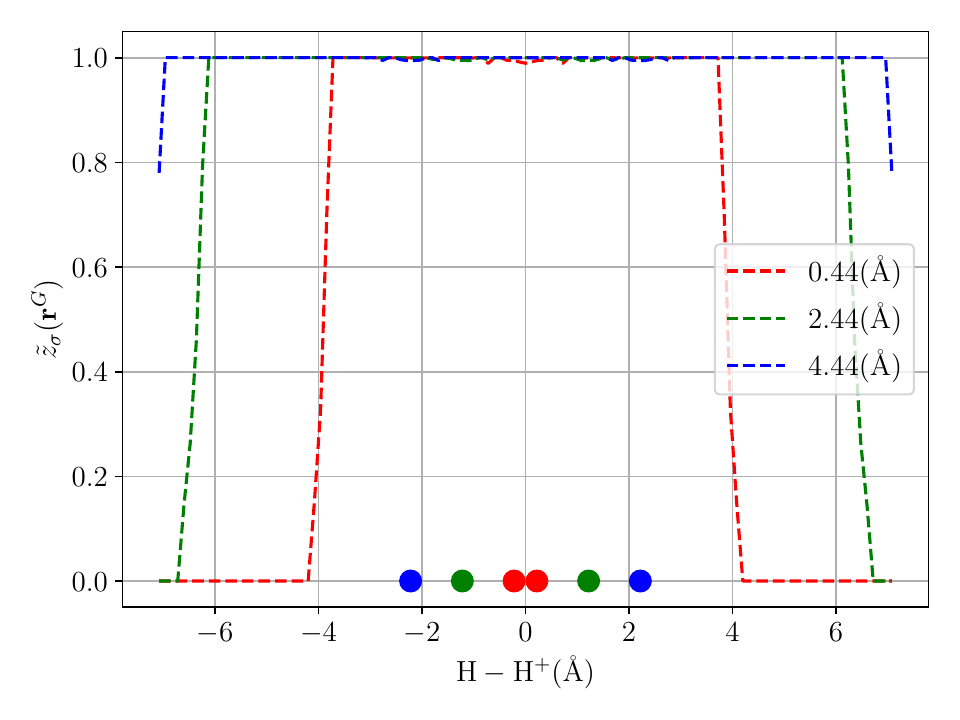}
\caption{Local scaling function evaluated 
with complex-valued pseudo-wavefunctions 
and pseudo-electron density of the 
H$^+_2$ molecule. The local scaling function 
is shown for different distances between 
the hydrogen and the hydrogen cation (indicated by the filled circles). 
In all cases the local scaling is approximately 1.0 
in all space where the electron density is above a set threshold of $n_\sigma(\bfr) > 1e-12$ [au].}
\label{fig:H2p_LS}
\end{figure}

In GPAW the ``all-electron'' electron 
spin density is given by
\begin{equation}
n_\sigma(\bfr) = \tilde{n}_\sigma(\bfr) + \sum_a(n_\sigma^a(\bfr) - \tilde{n}_\sigma^a(\bfr)) + \frac{1}{2}n^a_{c}(\bfr)
\end{equation}
where $\tilde{n}$ is a 
smooth pseudo valence density 
which matches exactly with the 
smooth atomic partial density 
$\tilde{n}^a$ within the PAW region. 
$n^a$ is an
``all-electron'' atomic partial density. 
Finally, $n^a_c$ are spherically symmetric 
atomic frozen core densities, 
where the factor $\frac{1}{2}$ accounts for the spin. 
We make use of the orthonormal property of the 
pseudo-valence states in UPAW 
and 
apply the locally scaled SIC energy functional 
to the pseudo density only. 
In this approximation the total 
pseudo-density is given by
\begin{align}
 \tilde{n}_\sigma &= \sum_k f_{k\sigma}\tilde{\psi}_{k\sigma}^*\tilde{\psi}_{k\sigma} + \frac{1}{2}\sum_a\tilde{\phi}^{a*}_c\tilde{\phi^a_c} \nonumber \\ 
 &= \sum_n f_{k\sigma}\tilde{\psi}_{k\sigma}^*\tilde{\psi}_{k\sigma} + \frac{1}{2}\sum_a\tilde{n}^a_c 
\end{align}
where the core electron density is 
given by smooth frozen core partial waves, 
$\tilde{\phi}^a_c$. 
The local scaling function in eq. ~\ref{eq:actual} becomes
\begin{equation}
 \tilde{z}_\sigma(\bfr) = \frac{\frac{|\nabla\left(\tilde{n}_\sigma(\bfr) + 1/2\tilde{n}_c(\bfr)\right)|^2}{8\left(\tilde{n}_\sigma(\bfr) + 1/2\tilde{n}_c(\bfr)\right)} + \frac{\tilde{\mathbf{J}}_\sigma(\bfr)\cdot \tilde{\mathbf{J}}_\sigma(\bfr)}{2\tilde{n}_\sigma(\bfr)}}{\frac{1}{2}\sum_i|\nabla \tilde{\psi}_{i\sigma}(\bfr)|^2 + \frac{1}{4}\sum_{a\alpha}|\nabla \tilde{\phi}^a_\alpha(\bfr)|^2} \nonumber. 
 \label{eq:LS_core}
\end{equation}

\begin{figure*}[!tbp]
    \centering
    \includegraphics[width=.49\textwidth]{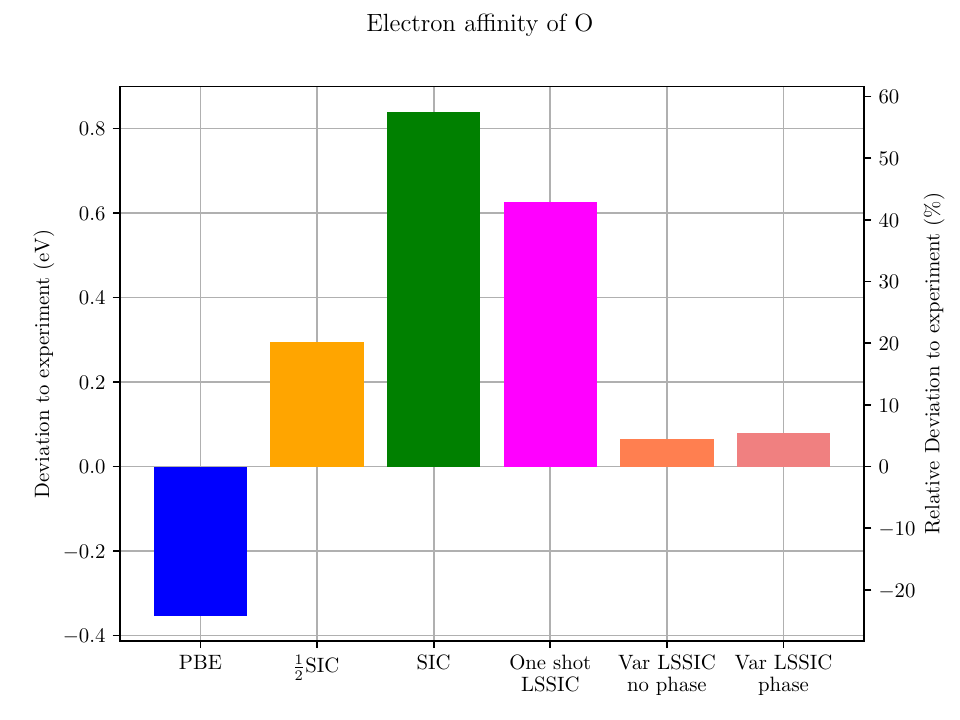}
    \includegraphics[width=.49\textwidth]{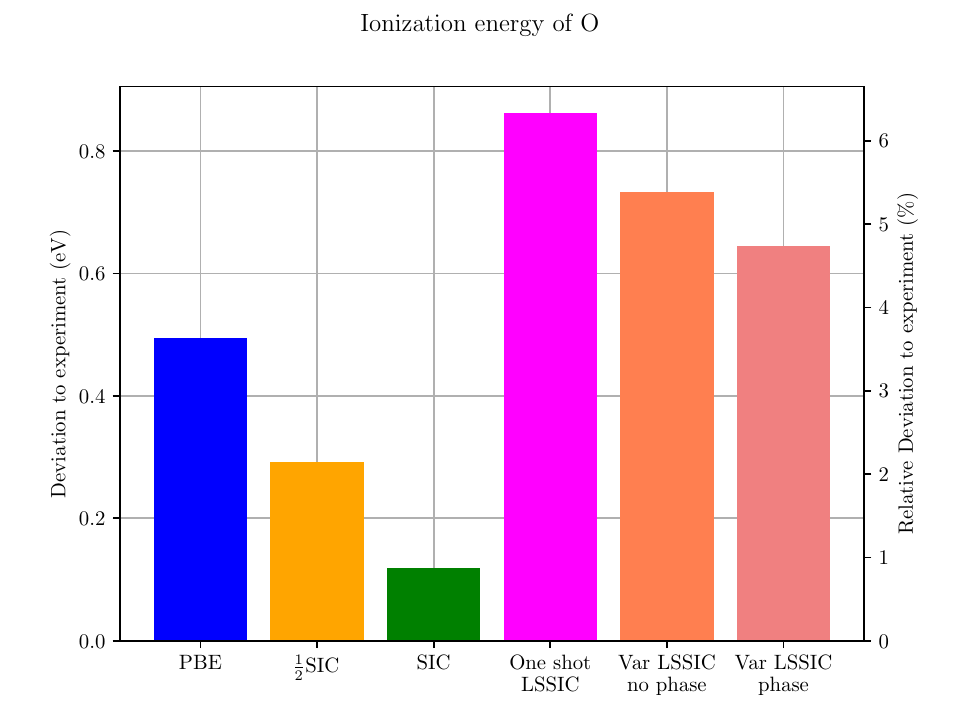}
    \caption{Left: bar plot showing the deviation of the calculated electron affinity from the experimental result. Right: bar plot showing the deviation of the calculated ionization energy from the experimental result. In both plots, the numerical values in electron-volts are given on the left y-axis, and the relative deviation is given on the right y-axis.}
    \label{fig:EAandIEofO}
\end{figure*}

The total self-interaction corrected energy functional becomes
\begin{align}
E^\mathrm{SIC} = &E^\mathrm{KS}[n] \nonumber \\
&- \sum_{i\sigma} \left( E_\mathrm{C}[\tilde{n}_{k\sigma},z_\sigma[\{\tilde{n}_{k\sigma}\}]] +  E_{xc}[\tilde{n}_{k\sigma}, 0, z_\sigma[\{\tilde{n}_{k\sigma}\}]]\right).
\label{eq:main}
\end{align}
A more detailed derivation 
of the local scaling function, 
resulting gradient and 
justification of the approximation given in eq.~\ref{eq:LS_core} 
is given in Appendix~\ref{Implementation} and~\ref{Derivation}. 
The gradient of the $\lss$ energy functional 
is evaluated numerically using finite-difference, 
and represents the first realization of 
a fully variational 
locally scaled 
self-interaction corrected energy functional 
based on complex optimal orbitals.

In all cases the 
KS-DFT exchange-correlation functional 
is PBE~\cite{PBE} and we adopt 
the following nomenclature: 
PBE, $\hsic+$PBE, SIC+PBE and $\lss$+PBE 
refer to calculations with the 
PBE energy functional, 
self-interaction corrected PBE energy functional 
scaled by $\hsic$ and 1 and 
locally scaled self-interaction corrected 
PBE energy functional using eq. \eqref{eq:actual}.
In some cases we also present results for $\lss$ with or 
without the complex phase term included, and "one shot" $\lss$, 
where we use an input pseudo-density and pseudo-wavefunctions 
from the PBE calculations
to evaluate the local scaling function -- and keep it fixed 
during the SCF calculation.

\section{Application}

In the following section the fully variational
$\lss$ 
using complex optimal orbitals are 
applied to simple example systems 
and compared to GGA calculations, 
using the PBE energy functional, 
as well as and half- and SIC calculations 
(i.e. energy functional of eq.~\eqref{eq:SIC} with $a=0.5$ and $a=1.0$).

\subsection{Hydrogen dimer cation}

We start with the simplest possible molecule, 
namely the hydrogen dimer cation, H$^+_2$, which has a 
single electron. It is well known that local 
and semi-local energy functionals predict the wrong 
dissociation limit where the electron is 
delocalized between the two hydrogens. 
The dimer binding potential energy surface is presented 
in Fig.~\ref{fig:H2p_BE} - and compared between 
the PBE energy functional, $\hsic$, SIC, 
and $\lss$ applied to the base functional. 
The hydrogen dimer cation is an important
test system since it represents the simplest
possible bond -- and contains a single electron. In this limit the PZ-SIC corrects exactly for the SIE -- and 
does indeed give the correct dissociation 
limit. In Table \ref{tbl:BEandBL}, the experimental bond energy and bond length of H$^+_2$ are compared to the PZ-SIC calculations. Although in terms of bond energy $\hsic$ yields the closest agreement to the experiment, it does not yield the correct dissociation limit since the binding energy is too low, because it only partially corrects for the SIE. SIC and variational LSSIC on the other hand do recover the correct dissociation limit, and at the same time provide a reasonable estimate for the bond energy and bond length of H$^+_2$. Unsurprisingly, these two calculations match exactly, since the localscaling function effectively scales the SIC by a factor of 1.0 in all space in the one-electron limit. H$^+_2$ described with the PBE energy functional 
dissociates into two partially charged hydrogen atoms, which yields an unphysical dissociation limit. Fig. ~\ref{fig:H2p_LS} shows the binding energy of H$_2^+$ and the dissociation limits of the PBE and SIC calculations.


\subsection{Electron affinity and ionization energy of atoms}

Table~\ref{tbl:EAandIE} lists the electron affinity (EA) 
and ionization energy (IE) of the Carbon(C), Nitrogen(N) and Oxygen(O) atoms, and Fig.~\ref{fig:EAandIEofO} 
presents bar plots showing the deviation of the calculated values from experiment for the O atom.
Results are presented for the PBE energy functional, $\hsic$, SIC and variational and one shot $\lss$ - and compared to experimental results 
(extracted from the \href{https://www.nist.gov}{NIST} database). 
In all cases the PBE functional gives an EA which is too low 
-- i.e. the additional electron is bound too strongly, and an IE 
which is too high -- i.e. it holds on the highest occupied 
valence electron too tightly.
$\hsic$ improves the estimate for the EA significantly, as 
well as for the IE, and gives the overall best result compared to experiment. SIC tends to overcorrect the SIE in the case of EA, 
and now instead the EA is too high compared to experiment 
-- i.e. the extra electron is bound to weakly. 
For the IE, SIC gives the best agreement with experiment, 
and deviates less than 1\% from experiment.
Variational $\lss$ predicts EAs for N and O in close agreement 
with experiments. They differences are 0.104eV and 0.065eV to the experimental results respectively. In fact, the variational $\lss$ calculations are closest to the experimental values compared to the other energy functionals. However, they underestimate the 
correction to the SIE in the case of the C EA. Variational $\lss$
only slightly improves the IE for C and N compared to the PBE
functional. Note that the effect of the inclusion of the 
complex phase term in the local scaling function
marginally improves the prediction of the IE for N and O,
as well as the EA, but is negligible in the EA and IE for C.
One shot $\lss$ performs the worst in comparison with the other SIE 
corrected energy functionals. It tends to over-correct the SIE when 
calculating the EA, and yields similar results when compared to SIC, whereas it completely 
under-corrects the SIE when calculating the IE, where it yields similar results as 
the base functional PBE for C and N.

All in all the variational $\lss$, including the gradient 
of the complex phase, gives the best agreement with experiment 
for the electron affinity of these atoms. For the ionization energies, it either only marginally
improves the results, or in the case of O
does not improve compared to the PBE functional. However, there is a
very clear improvement in the IE of O when using the $\lss$ 
with the gradient of the complex phase compared to the one-shot $\lss$ and $\lss$ 
without the gradient of the complex phase, this can be seen in Fig.~\ref{fig:EAandIEofO}.

\begin{table}[!tbp]
\centering
\caption{Equilibrium bond lengths and bond energies for C$_2$, N$_2$, O$_2$ and H$_2^+$. Results are shown for the PBE, $\hsic$ and SIC energy functionals, as well as $\lss$ with (phase) and without (no phase) the gradient of the complex phase included in the local scaling function. One shot $\lss$ is also presented where the local scaling function is calculated once with the converged PBE electron spin density. Experimental results were obtained from \cite{Darwent1970NSRDS31,NISTWebBook_SRD69}.}
\begin{tabular}{llcc}
\hline
Dimer & Calculation & Energy (eV) & Bond length ($\mathring{A}$) \\
\hline
\multirow{7}{*}{C$_2$}
 & PBE                     & -6.503 & 1.235 \\
 & $\frac{1}{2}$ SIC       & -6.140 & 1.217 \\
 & SIC                     & -5.879 & 1.206 \\
 & Var LSSIC no phase      & -6.466 & 1.211 \\
 & Var LSSIC phase         & -6.471 & 1.212 \\
 & Oneshot LSSIC no phase  & -6.380 & 1.210 \\
 & Experiment              & -6.355 & 1.242 \\
\hline
\multirow{7}{*}{N$_2$}
 & PBE                     & -10.514 & 1.104 \\
 & $\frac{1}{2}$ SIC       & -10.096 & 1.091 \\
 & SIC                     & -9.732  & 1.081 \\
 & Var LSSIC no phase      & -10.381 & 1.091 \\
 & Var LSSIC phase         & -10.382 & 1.091 \\
 & Oneshot LSSIC no phase  & -10.380 & 1.092 \\
 & Experiment              & -9.924  & 1.098 \\
\hline
\multirow{7}{*}{O$_2$}
 & PBE                     & -6.235 & 1.219 \\
 & $\frac{1}{2}$ SIC       & -5.479 & 1.188 \\
 & SIC                     & -4.840 & 1.166 \\
 & Var LSSIC no phase      & -5.400 & 1.190 \\
 & Var LSSIC phase         & -5.413 & 1.189 \\
 & Oneshot LSSIC no phase  & -5.201 & 1.190 \\
 & Experiment              & -5.213 & 1.207 \\
\hline
\multirow{5}{*}{H$_2^+$}
& PBE                     & -2.893 & 1.208 \\
& $\frac{1}{2}$ SIC       & -2.782 & 1.154 \\
& SIC                     & -2.694 & 1.117 \\
& Var LSSIC               & -2.694 & 1.117 \\
& Experiment              & -2.790 & 1.052 \\
\hline
\end{tabular}
\label{tbl:BEandBL}
\end{table}

\begin{table}[htbp]
\centering
\caption{Electron affinity (EA) and ionization energy (IE) for C, N, and O (eV). Results are shown for the PBE, $\hsic$ and SIC energy functionals, as well as $\lss$ with (phase) and without (no phase) the gradient of the complex phase included in the local scaling function. One shot $\lss$ is also presented where the local scaling function is calculated once with the converged PBE electron spin density. Experimental results were obtained from \cite{WolframResearch2007}.}
\begin{tabular}{llcc}
\hline
Element & Functional & EA (eV) & IE (eV) \\
\hline
\multirow{7}{*}{C} 
 & PBE                 & -1.606 & 11.622 \\
 & $\hsic$             & -1.290 & 11.450 \\
 & SIC               & -1.037 & 11.273 \\
 & Var LSSIC no phase  & -1.534 & 11.564 \\
 & Var LSSIC phase     & -1.535 & 11.563 \\
 & One shot SIC        & -1.007 & 11.505 \\
 & Experiment          & -1.261 & 11.26  \\
\hline
\multirow{7}{*}{N}
 & PBE                 & -0.331 & 14.917 \\
 & $\hsic$             &  0.251 & 14.773 \\
 & SIC               &  0.684 & 14.637 \\
 & Var LSSIC no phase  & -0.041 & 14.923 \\
 & Var LSSIC phase     & -0.033 & 14.858 \\
 & One shot SIC        &  0.427 & 14.739 \\
 & Experiment          &  0.071   & 14.53     \\
\hline
\multirow{7}{*}{O}
 & PBE                 & -1.813 & 14.104 \\
 & $\hsic$             & -1.165 & 13.902 \\
 & SIC               & -0.620 & 13.729 \\
 & Var LSSIC no phase  & -1.395 & 14.343 \\
 & Var LSSIC phase     & -1.381 & 14.254 \\
 & One shot SIC        & -0.833 & 14.472 \\
 & Experiment          & -1.460   & 13.61     \\
\hline
\end{tabular}
\label{tbl:EAandIE}
\end{table}

\begin{figure*}[!tbp]
    \centering
    \includegraphics[width=.99\textwidth]{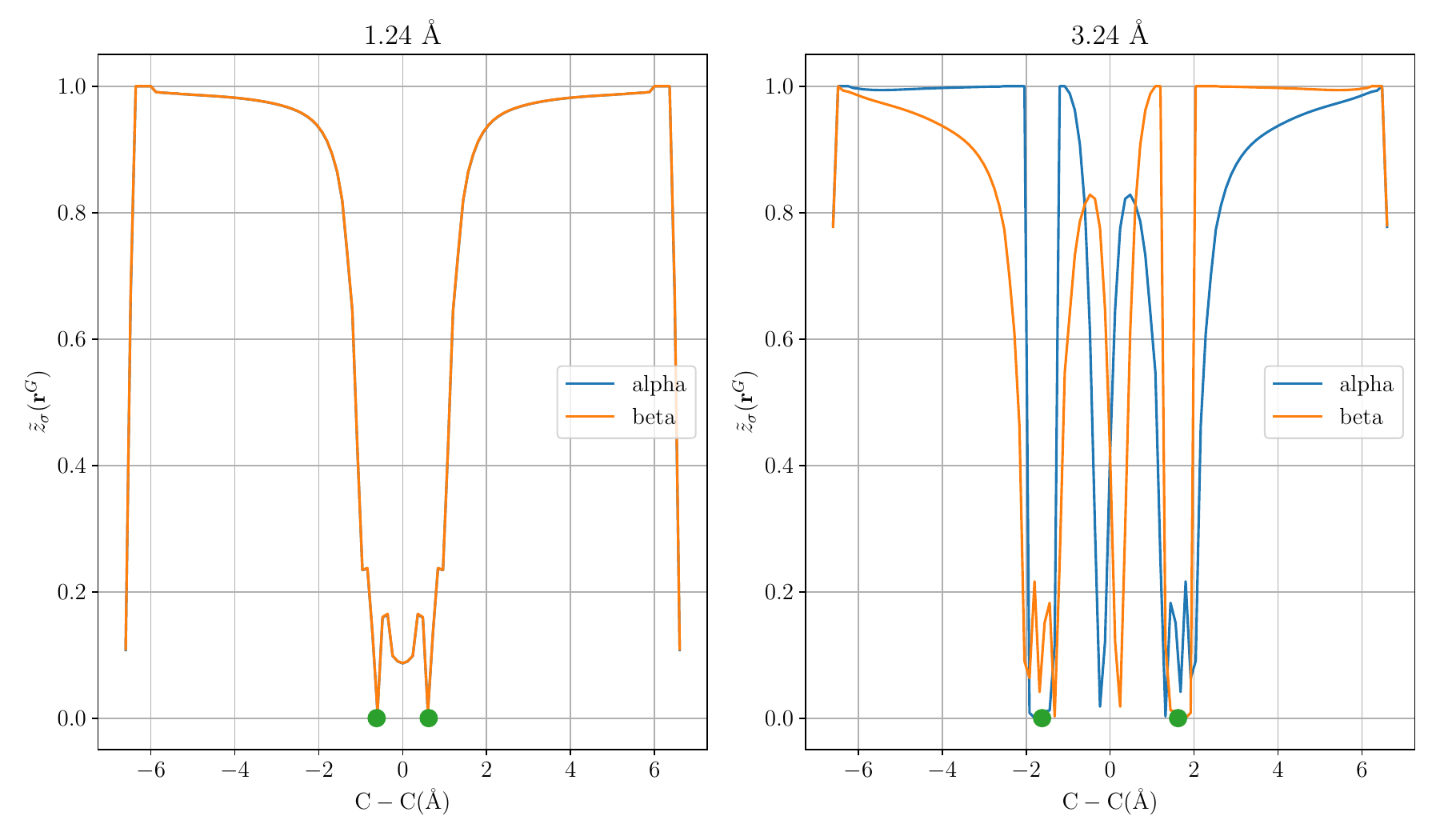}
    \caption{Localscaling function evaluated for the carbon dimer at two distances - 1.24 \AA, which is close to the experimental bond length, and 3.24 \AA. Both spin channels are presented and are denoted alpha (blue) and beta (orange). It is clear that the spin densities are symmetric in all space when the distance is close to the experimental bond length, whereas the spin densities are spatially different as we approach the dissociation limit, with one spin being preferentially conentrated on one carbon vs the other. The green dots denote the position of the carbon nuclei along the axis which is parallel to the bond.}
    \label{fig:LocalScalingC2}
\end{figure*}

\begin{figure}[!tbp]
    \centering
    \includegraphics[width=1.\linewidth]{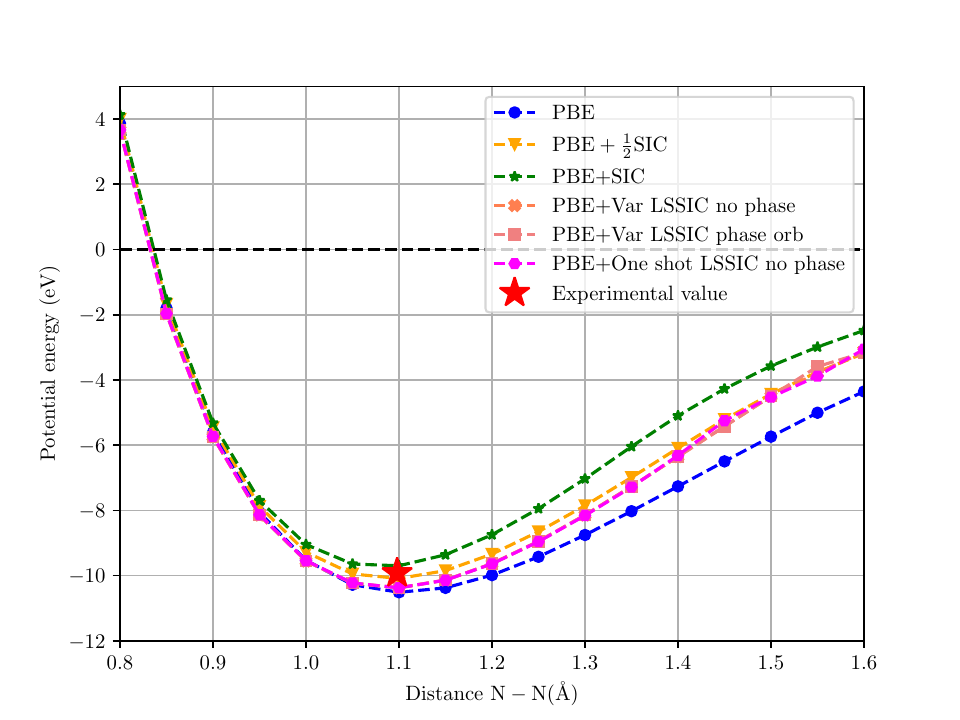}
    \caption{Potential energy of the nitrogen dimer as a function of distance. The experimental estimate is marked on the plot (red star).}
    \label{fig:PESN2}
\end{figure}

\subsection{Dimers}

Here the potential energy surface of the carbon, nitrogen and oxygen dimers are calculated using PBE, $\hsic$, SIC and the three flavors of $\lss$ - with and without the gradient of the phase term included, and one-shot $\lss$, where the input density is from a converged PBE calculation (and therefore has no phase since the converged PBE wavefunctions have no phase).
Fig.~\ref{fig:LocalScalingC2} presents the local scaling function for the carbon dimer evaluated along the bond axis and at two distances -- close to the experimental bond length, where the spin densities are spatially symmetric, and at a large separation where it is clear that one spin channel is localized preferentially on one of the carbon atoms (and vice versa). 

Table~\ref{tbl:BEandBL} presents the bond energy and bond lengths as predicted by all of the energy functionals and provides the corresponding experimental results. For C, N and O, PBE overestimates the bond energy -- with almost a 1 eV overestimation for the oxygen dimer bond energy -- but at the same time has the overall best agreement with the bond length compared to experiment. SIC on the other hand over corrects the SIE and predicts in all cases a bond energy which is too low, and a bond length which is too short. 

$\hsic$ presents a good compromise in all cases, since it on average yields the best agreement to experimental values in terms of both bond energy and bond length. $\lss$ with and without the gradient of the complex phase results in closer agreement to the experimental values in terms of the bond energy for both C and O compared to the PBE, $\hsic$ and SIC. One shot $\lss$ gives a remarkably good agreement for the bond energy of both C and O dimers. The bond energy calculated by One shot $\lss$ differs 0.025eV and 0.012eV from the experiment respectively. 

The greatest deviation for $\lss$ in terms of the bond energy is for the N dimer, and the various different potential energy surfaces are shown in Fig.~\ref{fig:PESN2}. $\lss$ only slightly improves the bond energy compared to PBE in this case. While $\lss$ predicts a too large bond energy, compared to experiment and $\hsic$, the general curve after the equilibrium bond length tends to follow the $\hsic$ curve.





\section{Conclusion \& Outlook}

We have presented a fully variational locally scaled self-interaction corrected energy functional, using complex optimal orbitals. The new energy functional captures the exact PES of the hydrogen cation - since in the limit of a single electron the SIC energy is captured (and is formally an exact correction). In the case of the C, N and O atoms the locally scaled SIC improves the prediction of electron affinity, whereas it does not improve the prediction of ionization energies, compared to the PBE functional. For the multi-electron dimers considered in this work, all LSSIC energy functionals predict a reasonable bond energy and equilibrium bond distance, i.e. within 4\% and 3\% of the experimental value respectively. This shows that the local scaling SIC energy functional is either on a similar footing as other energy functional methods, or improve the predicted properties. 
The local scaling SIC is a general framework, in the sense that the local scaling function can be tweaked and tuned by introducing scaling parameters and or other functions based on the electron density. In future work the shape of the local scaling function will be optimized in order to best capture the properties of atoms and molecules, as well as defect states in solid-systems.

\begin{acknowledgements}
We thank Prof. Gianluca Levi and Dr. Aleksei Ivanov for fruitful discussions and insights. 
This work was supported by the Innovation Fund, grant agreement no.\ 2513106, and the Icelandic Research Fund, grant agreement no.\ 2410644. Computer resources, data storage, and IT user support were provided by the 
Icelandic Research e-Infrastructure (IREI), funded by the Icelandic Infrastructure Fund.
\end{acknowledgements}

\bibliography{References}

\appendix

\section{Implementation in GPAW}
\label{Implementation}

Within the PAW formalism, the so-called ``all-electron'' wave functions, which contain cusps at the positions of the nuclei, are written as
\begin{equation}
 \psi_{i,\sigma}(\mathbf{r}) = \hat{\mathcal{T}}\tilde{\psi}_{i,\sigma}(\mathbf{r})
\end{equation}
where $\tilde{\psi}_{i,\sigma}$ are ``pseudo-electron'' wave functions, which are smooth everywhere. $\hat{\mathcal{T}}$ is a linear transformation operator, which corrects for the smooth description of the electronic wave functions near the positions of the nuclei 
\begin{equation}
 \hat{\mathcal{T}} = 1 + \sum_{a\alpha}(\ket{\varphi^a_\alpha} - \ket{\tilde{\varphi}^a_\alpha})\bra{\tilde{p}^a_\alpha}
\end{equation}
Here, $\varphi^a_\alpha$ and $\tilde{\varphi}^a_\alpha$ are partial waves describing the all-electron and pseudo-electron wave functions in an atomic region of radius $r^a_c$ around each nucleus $a$.
The all-electron and pseudo-electron partial waves are required to be identical beyond the radius $r^a_c$, i.e. $\varphi^a_\alpha(\mathbf{r})=\tilde{\varphi}^a_\alpha(\mathbf{r})$ for $\mid\mathbf{r} - \mathbf{R}^a\mid > r^a_c$.
$\tilde{p}^a_\alpha$ are smooth projection functions, which satisfy 
\begin{align}
 \sum_\alpha \ket{\tilde{\varphi}^a_\alpha} \bra{\tilde{p}^a_\alpha} =&\, 1 \ \ \ \ \text{for} \mid\mathbf{r} - \mathbf{R}^a\mid \leq r^a_c \\
 \braket{\tilde{p}^a_\alpha| \tilde{\phi}^a_\beta} =&\, \delta_{\alpha\beta} \ \ \ \ \text{for} \mid\mathbf{r} - \mathbf{R}^a\mid \leq r^a_c
\end{align}
such that
\begin{align}
 \ket{\tilde{\psi}_{i,\sigma}} =& \sum_\alpha \ket{\tilde{\varphi}^a_\alpha} P^a_{\alpha i,\sigma} \ \ \ \ \text{for} \mid\mathbf{r} - \mathbf{R}^a\mid \leq r^a_c \\
 \ket{\psi_{i,\sigma}} =& \sum_\alpha\ket{\varphi^a_\alpha} P^a_{\alpha i,\sigma} \ \ \ \ \text{for} \mid\mathbf{r} - \mathbf{R}^a\mid \leq r^a_c
\end{align}
where $P^a_{\alpha i,\sigma} = \braket{\tilde{p}^a_\alpha| \tilde{\psi}_{i,\sigma}}$, i.e. the all- and pseudo-electron wave functions can be expanded into partial waves with the same linear expansion coefficients.

We use unitary projector augmented wave (UPAW), which has recently been implemented in GPAW~\cite{ivanov2024UPAW} to describe the frozen core electrons and pseudo- to all-electron transformation. In UPAW the transformation operator is enforced to fulfill:
\begin{equation}
 \hat{O} = \hat{\tau}^\dagger\hat{\tau} = I
\end{equation}
i.e. $\hat{\tau}^\dagger = \hat{\tau}^{-1}$. This is realized by enforcing the partial wave projector operator to satisfy:
\begin{equation}
\Delta \hat{O} = \sum_a\sum_{ij}\ket{\tilde{p}_i^{a}}O^a_{ij}\bra{\tilde{p}_j^a} = 0
\end{equation}
which is guaranteed if
\begin{equation}
 O^a_{ij} = \langle \phi^a_i|\phi^a_j\rangle - \langle \tilde{\phi}^a_i|\tilde{\phi}^a_j\rangle = 0
\end{equation}
Hence, in order to construct a UPAW augmentation region the overlap of the $ij$ pair of pseudo partial waves and the $ij$ pair of all-electron partial waves must be the same. A direct consequence of UPAW is that the pseudo-valence states naturally become orthonormal
\begin{equation}
\int_V\tilde{\psi}^*_n(\mathbf{r})\tilde{\psi}_m(\mathbf{r})d\mathbf{r} = \delta_{nm}
\end{equation}
and the pseudo-valence density integrates to the stoichiometric number of valence electrons (in charge neutral systems), or
\begin{equation}
 \int_V\tilde{n}(\mathbf{r})d\mathbf{r} = \int_Vn(\mathbf{r})d\mathbf{r} = n_\mathrm{valence}
\end{equation}
The Hamiltonian fully separates into a pseudo and atomic PAW correction, and the pseudo part can be trivially solved
\begin{equation}
 (\tilde{\hat{H}}(\mathbf{r}) - \epsilon_n)\tilde{\psi}_n(\mathbf{r}) = 0
\end{equation}
i.e. the pseudo-valence states are eigenvectors of the pseudo-Hamiltonian. Atomic corrections can be selectively included in the Hamiltonian without loss of generality, due to the linear mapping between the atomic and grid region.

Therefore we write an approximate, but variationally self-consistent, SIC term with scaling as
\begin{align}
E^\mathrm{SIC} = &E^\mathrm{KS}[n] \nonumber \\
&- \frac{1}{2}\sum_{i\sigma} \left( E_\mathrm{C}[\tilde{n}_{i\sigma},z[\tilde{n}_{i\sigma}]] + E_{xc}[\tilde{n}_{i\sigma}, z[\tilde{n}_{i\sigma}], 0]\right)
\end{align}
i.e. we do not calculate the SIC part of the PAW correction associated with the UPAW. 

The energy functional and resulting Hamiltonian is orbital density dependent and therefore 
requires
an optimization method which finds the optimal orbitals that minimizes the total SIC energy functional
$$\underset{{\psi_i}}{\mathrm{min}} \ E^\mathrm{SIC}[\{\psi_i\}]$$
To that end the minimum of the functional is found via a direct optimization method where the optimal orbitals are found variationally and self-consistently \cite{Ivanov_Jonsson_Vegge_Jonsson_2021,ivanov2021levijonssonjonsson,Schmerwitz_Ivanov_Jonsson_Jonsson_Levi_2022}.
These orbitals must satisfy the Euler-Lagrange equations (the spin index $\sigma$ is omitted for clarity):
\begin{align}\label{eq: KSequation}
    & f_i\left(
    \hat h  + \hat v_i \right) \ket{\psi_{i}} = \sum_{j} \ket{\psi_j} \lambda_{ji} \\
   \label{eq: KSpotential}
   & \hat h = 
    -\frac{1}{2} \Delta + v_{ext} + v_{C}[n] + v_{xc}[n] \\
    &\hat v_i = -\frac{1}{2}\left( v_{C}[\tilde{n}_i,z[\tilde{n}_i]] + v_{xc}[\tilde{n}_i,z[\tilde{n}_i]] \right)
\end{align}
where $f_i$ is the occupation number of the  $i$\textsuperscript{th} orbital $\ket{\psi_{i}}$,  $\lambda_{ji}$ is a Lagrange multiplier which enforces orthogonality between orbital pairs $ij$. $v_{ext}$ is the external potential, $v_{H}$ and $v_{xc}$ are Hartree and the exchange-correlation potentials, respectively,  $n$ is the total density, and $n_i$ is the $i$\textsuperscript{th} orbital density.

Due to the local scaling function additional terms are needed (other than the orbital dependent Coulomb and exchange-correlation potential) in the orbital dependent potential operator $\hat v_i$. These are

\begin{align}
v_C[\tilde{n}_i&,z[\tilde{n}_i]] \nonumber \\
&= \sum_j\frac{\partial E_C[\tilde{n}_{i},z[\tilde{n}_i]]}{\partial \tilde{n}_{j}(\bfr'')} \nonumber \\
&= \frac{1}{2}\sum_j \frac{\partial}{\partial \tilde{n}_j(\bfr'')} \iint \frac{z(\bfr) \tilde{n}_i(\bfr) \tilde{n}_i(\bfr')}{|\bfr-\bfr'|} d\bfr d\bfr' \nonumber \\
&= \frac{1}{2}z(\bfr'') \int\frac{\tilde{n}_i(\bfr')}{|\bfr'-\bfr''|} d\bfr' + \frac{1}{2}\int\frac{z(\bfr)\tilde{n}_i(\bfr)}{|\bfr-\bfr''|} d\bfr 
\nonumber \\
&\ \ \ \ + \frac{1}{2}\sum_j \frac{\partial z(\bfr'')}{\partial \tilde{n}_i(\bfr'')}\tilde{n}_j(\bfr'') \int\frac{\tilde{n}_j(\bfr')}{|\bfr''-\bfr'|} d\bfr'
\nonumber \\
\end{align}
and
\begin{align}
v_{xc}[\tilde{n}_i&,z[\tilde{n}_i]] \nonumber \\
=& \sum_j\frac{\partial E_{xc}[\tilde{n}_i,0,z[\tilde{n}_i]]}{\partial \tilde{n}_j(\bfr'')} \nonumber \\
=& \sum_j \frac{\partial}{\partial \tilde{n}_j(\bfr'')}\int z(\bfr)\tilde{n}_i(\bfr)\epsilon_{xc}(\tilde{n}_i,0)d\bfr \nonumber \\
=& z(\bfr'')\tilde{v}_{xc,i}(\bfr'') \nonumber \\
&+ \sum_j\frac{\partial z(\bfr'')}{\partial \tilde{n}_i(\bfr'')}\tilde{n}_j(\bfr'')\epsilon_{xc}(\tilde{n}_j,0)
\end{align}


\section{Derivation of the local scaling function}
\label{Derivation}
The following provides a concise derivation of the kinetic energy densities entering the iso-orbital indicator used as a local scaling function in PZ-SIC. The derivation closely follows that of Romanowski and Krukowski \cite{Romanowski2009}, but is written here allowing for complex-valued orbitals. The resulting expressions define the kinetic energy densities used in this work, which are subsequently combined to form the local scaling function.

\subsubsection*{Generalised von Weizsacker Kinetic Energy Density}
We start with the standard definition of kinetic energy T for a single wavefunction  $\psi(\mathbf{r})$:
\begin{equation}
T = -\frac{1}{2} \int_{\mathbb{R}^3} \psi^*(\mathbf{r}) \nabla^2 \psi(\mathbf{r}) \, d\mathbf{r}.
\label{T_definition}
\end{equation}

and electron density, defined as: 
\begin{equation}
    n(\mathbf{r})
= \psi(\mathbf{r})^* \psi(\mathbf{r}).
\end{equation}
Taking the Laplacian on both sides:

\begin{equation}
\nabla^2 n(\mathbf{r}) 
= 2 \Re\left\{ \psi(\mathbf r)^* \nabla^2 \psi(\mathbf r) \right\} + |\nabla \psi(\mathbf r)|^2 
\end{equation}

and integrating over $\mathbb{R}^3$ gives

\begin{equation}
\begin{aligned}
\int_{\mathbb{R}^3} \nabla^2 n(\mathbf{r}) \, d\mathbf{r}
&= \int_{\mathbb{R}^3} 2\Re\!\left\{ \psi^*(\mathbf{r}) \nabla^2 \psi(\mathbf{r}) \right\} \, d\mathbf{r} \\
&\quad + 2 \int_{\mathbb{R}^3} |\nabla \psi(\mathbf{r})|^2 \, d\mathbf{r} \\
&=-4\, \left( -\frac{1}{2} \int_{\mathbb{R}^3} \psi^*(\mathbf{r}) \nabla^2 \psi(\mathbf{r}) \, d\mathbf{r} \right) \\
&\quad + 2 \int_{\mathbb{R}^3} |\nabla \psi(\mathbf{r})|^2 \, d\mathbf{r} \\
&= -4T + 2 \int_{\mathbb{R}^3} |\nabla \psi(\mathbf{r})|^2 \, d\mathbf{r}.
\end{aligned}
\end{equation}

Rearranging, we obtain:
\begin{equation}
T = \int_{\mathbb{R}^3} \left[ \frac{1}{2} |\nabla \psi(\mathbf{r})|^2 - \frac{1}{4} \nabla^2 n(\mathbf{r}) \right] \, d\mathbf{r}.
\label{T_T}
\end{equation}

Let us now assume a complex wave function
\begin{equation}
\psi(\mathbf{r}) = \sqrt{n(\mathbf{r})} e^{i\theta(\mathbf{r})},
\label{complex_wf}
\end{equation}
where $\theta(\mathbf{r})$ is the local phase.

In this form:
\begin{align}
\label{complex_wf_laplacian}
\nabla \psi(\mathbf r) &= \frac{\nabla n(\mathbf r)}{2\sqrt{n(\mathbf r)}} \, e^{i\theta(\mathbf r)}
+ i \sqrt{n(\mathbf r)} \, \nabla \theta(\mathbf r) \, e^{i\theta(\mathbf r)}, \\
|\nabla \psi(\mathbf r)|^2 &= \frac{|\nabla n(\mathbf r)|^2}{4\,n(\mathbf r)}
+ n(\mathbf r)\,|\nabla \theta(\mathbf r)|^2 .
\label{eq:laplacian_madelung_mod_squared}
\end{align}

Substituting \eqref{eq:laplacian_madelung_mod_squared} into \eqref{T_T} the integrand becomes
\begin{equation}
\tau^{\text{g-vW}}(\mathbf{r}) = \frac{|\nabla n(\mathbf{r})|^2}{8n(\mathbf{r})} + \frac{1}{2} n(\mathbf{r}) |\nabla \theta(\mathbf{r})|^2 - \frac{1}{4} \nabla^2 n(\mathbf{r}).
\end{equation}

\subsubsection*{Generalised Kohn--Sham Kinetic Energy Density}

We extend the kinetic energy decomposition to a system of $N$ orthonormal Kohn--Sham orbitals $\psi_i(\mathbf r)$. The total kinetic energy is
\begin{equation}
T = -\frac{1}{2} \sum_{i=1}^N \int_{\mathbb{R}^3} \psi_i^*(\mathbf r)\,\nabla^2\psi_i(\mathbf r)\,d\mathbf r .
\end{equation}
The total electron density is defined as
\begin{equation}
n(\mathbf r) = \sum_{i=1}^N |\psi_i(\mathbf r)|^2 .
\end{equation}
Proceeding as in the single-orbital case and using the linearity of sums, integrals, and the Laplacian, we obtain
\begin{equation}
T = \int_{\mathbb{R}^3}
\left[
\sum_{i=1}^N \frac{1}{2} |\nabla \psi_i(\mathbf r)|^2
- \frac{1}{4} \nabla^2 n(\mathbf r)
\right]
d\mathbf r .
\end{equation}
This naturally defines the generalised Kohn--Sham kinetic energy density
\begin{equation}
\tau^{\mathrm{g\text{-}KS}}(\mathbf r)
=
\sum_{i=1}^N \frac{1}{2} |\nabla \psi_i(\mathbf r)|^2
- \frac{1}{4} \nabla^2 n(\mathbf r) .
\end{equation}

\subsubsection*{Complex Phase Term in Kohn-Sham DFT}
For a system of $N$ orthonormal Kohn-Sham orbitals the gradient of the phase term can be cast in terms of the current density, which is given by:
$$\mathbf{J}(\mathbf{r}) = -\frac{i}{2}\sum_k^{\mathrm{occ}}\left[\psi^*_k(\mathbf{r})\nabla \psi_k(\mathbf{r}) - \psi_k(\mathbf{r})\nabla \psi^*_k(\mathbf{r})\right]$$

The gradient of the phase $|\nabla\theta(\mathbf{r})|^2$ can be related to the current density computing:

\begin{equation}
\begin{aligned}
\mathbf{J}_\sigma(\mathbf r)\cdot\mathbf{J}_\sigma(\mathbf r)
&= -\frac{1}{4}
\sum_{k}^{\mathrm{occ}}
\Big[
\psi_{k\sigma}^*(\mathbf r)\,\nabla\psi_{k\sigma}(\mathbf r)
-
\psi_{k\sigma}(\mathbf r)\,\nabla\psi_{k\sigma}^*(\mathbf r)
\Big]^2
\\[4pt]
&= -\frac{1}{4}
\Big[
\big(
\psi_{k\sigma}^*(\mathbf r)\,\nabla\psi_{k\sigma}(\mathbf r)
\big)^2
-
\big(
\psi_{k\sigma}(\mathbf r)\,\nabla\psi_{k\sigma}^*(\mathbf r)
\big)^2
\\
&\quad
-|\psi_{k\sigma}(\mathbf r)|^2
\,|\nabla\psi_{k\sigma}(\mathbf r)|^2
\Big]
\end{aligned}
\end{equation}

Assuming complex wavefunctions \eqref{complex_wf}, \eqref{complex_wf_laplacian}, combining and canceling terms, this dot product yields:

\begin{equation}
\begin{aligned}
\mathbf{J}_\sigma(\mathbf r)\cdot\mathbf{J}_\sigma(\mathbf r)
&= 
\sum_{k}^{\mathrm{occ}}
n_{k\sigma}^2(\mathbf r)|\nabla\theta_{k\sigma}(\mathbf{r})|^2.
\end{aligned}
\end{equation}

which can be related to the gradient of the phase term~\cite{becke1996current,johnson2007density} via 
$$\frac{1}{2}n_{\sigma}(\mathbf{r})|\nabla\theta(\mathbf{r})|^2 = \frac{\mathbf{J(\mathbf{r})\cdot J(\mathbf{r})}}{2n_\sigma(\mathbf{r})}.$$

\subsubsection*{Scaling function}

Combining the generalised von Weizs\"acker and Kohn--Sham kinetic energy densities,
we obtain the local scaling function
\begin{equation}
z_\sigma(\mathbf r)
=
\frac{
\frac{|\nabla n_\sigma(\mathbf r)|^2}{8 n_\sigma(\mathbf r)}
+ \frac{1}{2} n_\sigma(\mathbf r)\,|\nabla \theta_\sigma(\mathbf r)|^2
- \frac{1}{4} \nabla^2 n_\sigma(\mathbf r)
}{
\frac{1}{2} \sum_i f_{i\sigma} |\nabla \psi_{i\sigma}(\mathbf r)|^2
- \frac{1}{4} \nabla^2 n_\sigma(\mathbf r)
}
\end{equation}

where $\sigma$ denotes a spin channel and $f_{i\sigma}$ the corresponding occupation numbers.



\end{document}